\documentclass[]{spie}  

\usepackage{amsmath,amsfonts,amssymb}
\usepackage{graphicx}
\usepackage[colorlinks=true, allcolors=blue]{hyperref}
\usepackage{units}
\usepackage{tikz}

\graphicspath{{figures/}}

\title{First Tests of the I-BEAT Detector as Primary Monitor for Target Normal Sheath Accelerated Protons}

\author[a]{F. Balling}
\author[a]{S. Gerlach}
\author[a]{A.-K. Schmidt}
\author[b]{V. Bagnoud}
\author[b, c, d]{J. Hornung}
\author[b]{B. Zielbauer}
\author[a]{K. Parodi}
\author[a]{J. Schreiber}
\affil[a]{LMU Munich, Am Coulombwall 1, 85748 Garching, Germany}
\affil[b]{GSI Helmholtzzentrum f\"ur Schwerionenforschung GmbH, Planckstr. 1, 64291 Darmstadt, Germany}
\affil[c]{Friedrich-Schiller-Universit\"at Jena, F\"urstengraben 1, 07743 Jena, Germany}
\affil[d]{Helmholtz-Institut Jena, Fröbelstieg 3, 07743 Jena, Germany}

\authorinfo{Corresponding author: Felix Balling, Felix.Balling@physik.lmu.de}

\begin{document} 
\maketitle

\begin{abstract}
The properties of laser-accelerated ion bunches are demanding and require development of suitable beam diagnostics. In particular, the short and intense particle bunches with a broad energy spectrum emitted in conjunction with a strong electromagnetic pulse (EMP) are challenging for conventional and well established monitoring systems. An approach based on measuring the acoustic signals of particles depositing their energy in water, referred to as ionoacoustics \cite{doi:10.1142/S0217732315400258}\cite{https://doi.org/10.1118/1.4905047} was recently developed into Ion-Bunch Energy Acoustic Tracing (I-BEAT)\cite{haffa_i-beat:_2019}. I-BEAT allows online detection of single proton bunches while being cost effective and EMP resistant. A simple water phantom equipped with only one ultrasound transducer positioned on the ion axis allows for reconstructing a rather complex energy spectrum that is typical for (manipulated) laser-accelerated ion bunches. To deduce the lateral bunch properties, additional transducers can be added, for example perpendicular to the ion beam axis.

This established setup has been adapted for use closely behind the laser target and tested at the PHELIX laser at GSI. The capability of the system to retrieve information about the broad proton spectrum close to the source despite the harsh conditions has been demonstrated. Future improvements are required, most importantly the increase of dynamic range. Nevertheless, I-BEAT holds promise to evolve into an online diagnostic tool particularly suited for laser-driven source development and optimization at high repetition rates.

\end{abstract}


\section{INTRODUCTION}
\label{sec:intro}

Over the last decades development of laser-driven ion sources\cite{daido2012review} and their applications have progressed towards higher proton energies\cite{higginson2018near} and radio-biological experiments.\cite{Kraft_2010}\cite{doi:10.1063/5.0008512}
One important asset for optimisation and application are instruments tailored to the particular characteristics of laser-driven ion bunches. Many solutions currently offer absolute dosimetry but rely on offline methods (e.g. radiochromic films, imaging plates, CR39 films). This poses restrictions, in particular as the repetition rates of PW-class systems have reached the 1-Hz level. For source development at these repetition rates we are in need of a detector that provides immediate feedback (ideally dosimetrically like films) and can operate in harsh environment close to the ion source. It must therefore resist high doses and dose rates as well as the strong electromagnetic pulse (EMP) emitted by the laser-plasma interaction, and it must promise a high dynamics range in principle.

We have proposed the I-BEAT technique detector\cite{haffa_i-beat:_2019} to  mitigate these problems of online detectors by stopping the ions in a water reservoir and measuring the ultrasonic acoustic wave originating from the energy deposition volume. \cite{doi:10.1142/S0217732315400258}\cite{https://doi.org/10.1118/1.4905047} Due to the low speed of sound in water, the detected acoustic signal is temporally separated from the EMP. As the detection medium is liquid water it is not damaged by radiation and can be exchanged when activated. In its first demonstration, we used only one transducer that provided a measurement along the beam entrance axis and we employed I-BEAT only in the final focus of a refocused and hence energy-selected proton bunch. This had simplified matters, as the depth-dose profile (and thus the source of the pressure pulse) was well confined, leading to a simple pulse structure. This is not expected for the case directly behind the target, where the energy and angular distribution is broad.
In this proceeding, we report on first experimental results with I-BEAT operated a few cm behind the primary laser-proton source (i.e. the target), without energy selection. We observe pressure signals that are clearly related to the impact of intense proton bunches. In this sense, these first tests are considered successful. In its current state, it remains difficult though to extract absolute information other than a mean penetration depth or an overall signal strength for a particular shot. This is due to the complicated signal structure that still requires a more in-depth analysis. Future investigations must therefore clarify the origin of the observed modulations and we must increase the dynamic range of the employed pressure measurement.

\section{EXPERIMENTS}
\label{sec:exp}

\begin{figure} [ht]
  \begin{center}
  \begin{tabular}{cc} 
   \begin{tikzpicture}
   \node[anchor=south west, inner sep=0](X) at %
   (0, 0){\includegraphics[width=.4\textwidth]{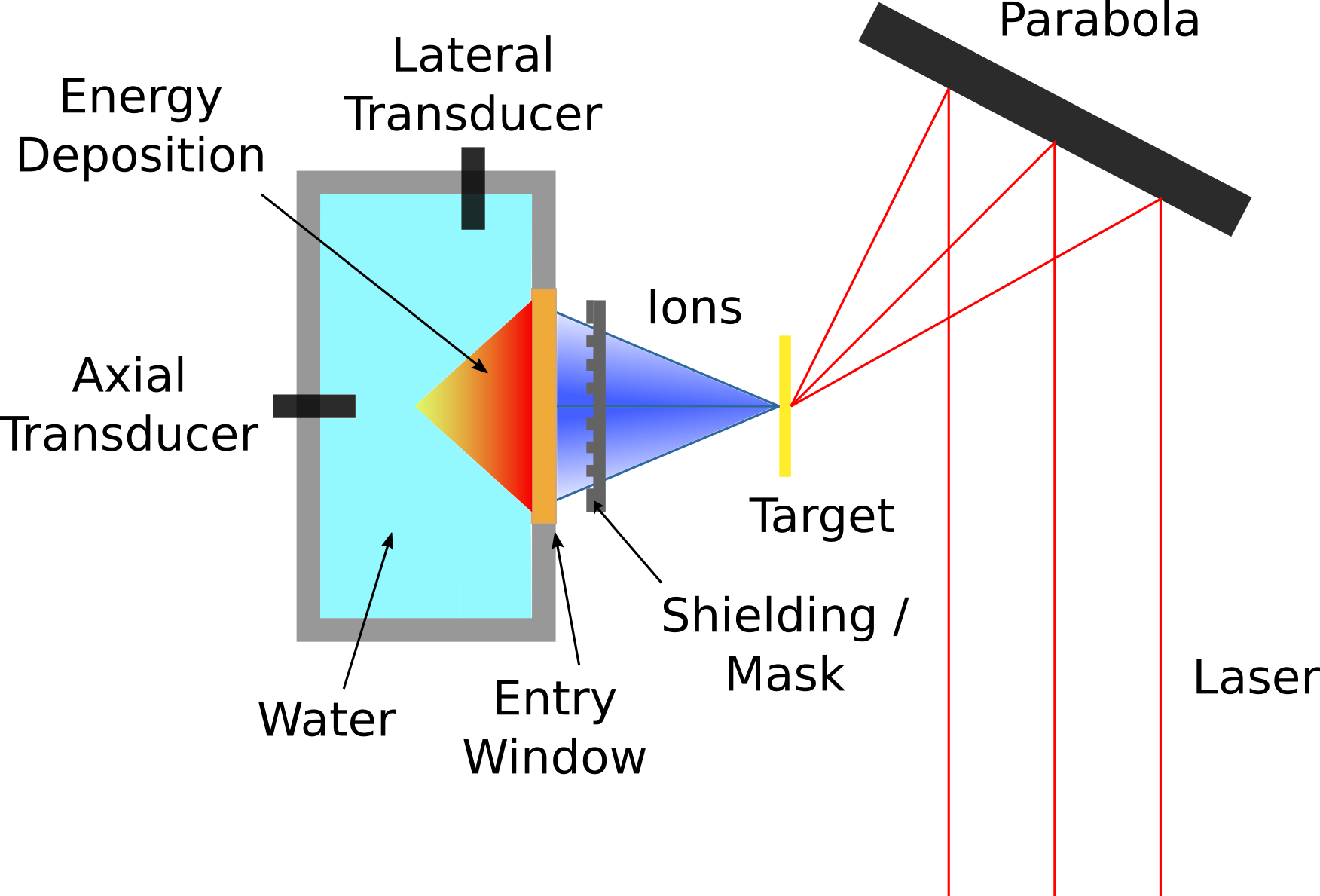}};%
   \begin{scope}[x={(X.south east)}, y={(X.north west)}]%
   \node[anchor=north west, black, fill=white](Y) at (0, 1.1){a)};%
   \end{scope}
   \end{tikzpicture} &
  \begin{tikzpicture}
   \node[anchor=south west, inner sep=0](X) at %
   (0, 0){\includegraphics[width=.4\textwidth]{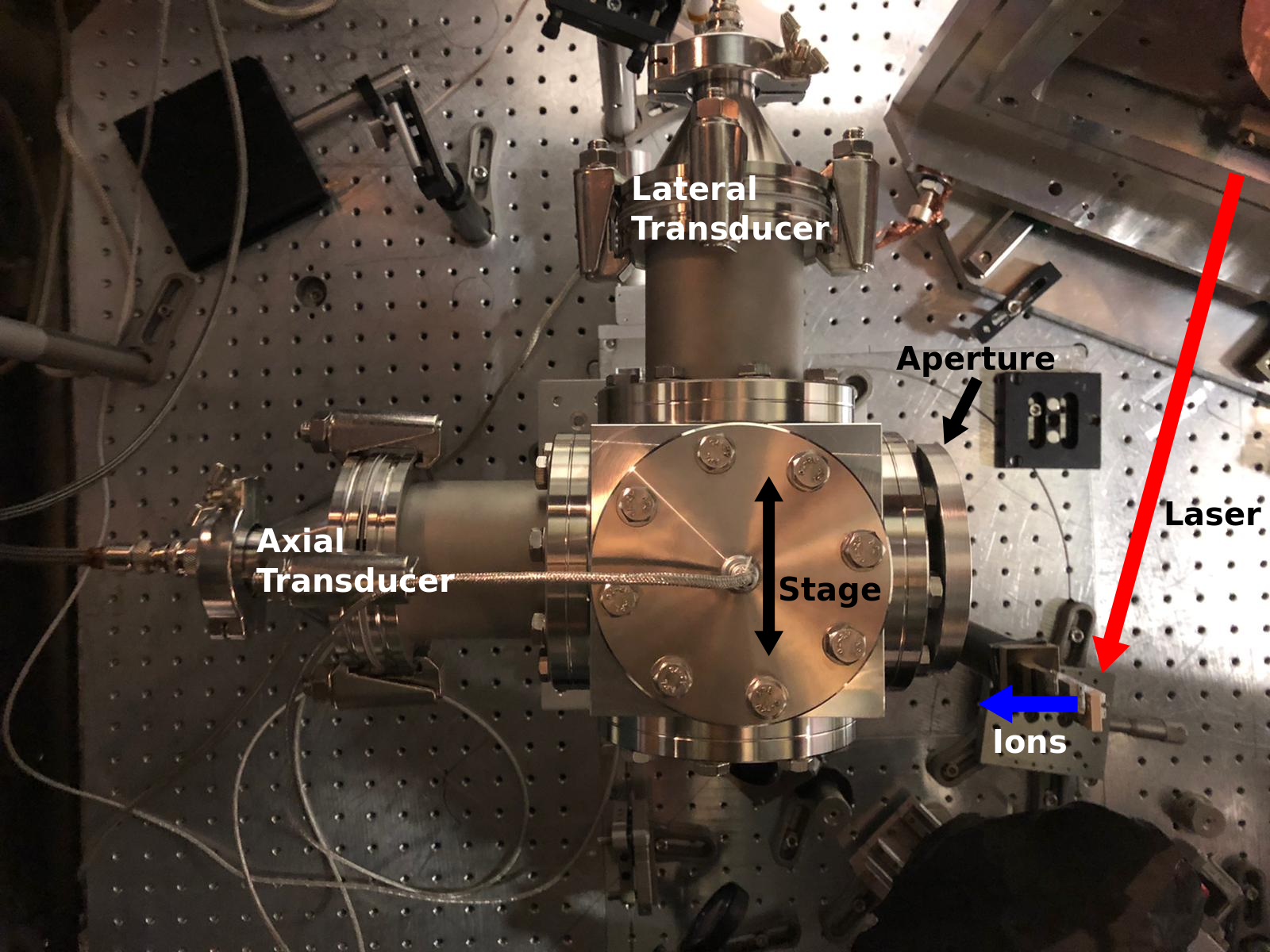}};%
   \begin{scope}[x={(X.south east)}, y={(X.north west)}]%
   \node[anchor=north west, black, fill=white](Y) at (0, 1){b)};%
   \end{scope}
   \end{tikzpicture}
  \end{tabular}
  \end{center}
  \caption[Exp-Setup] 
  { \label{fig:exp-setup} 
The PHELIX laser pulse is focussed by an F/1.4 OAP onto the \unit[1]{\textmu m} thick target. The ion bunch is modulated spatially by an Aluminum mask before entering the water reservoir through a \unit[125]{\textmu m} Kapton window. The ultrasonic sound wave can in principle be detected by two transducers: one on the ion axis ('Axial Transducer') and one perpendicular to the ion axis ('Lateral Transducer'). For comparison on a single-shot basis, it is possible to immerse radiochromic films (RCFs) in the reservoir. The first film is placed \unit[10]{mm} from the entrance window with \unit[8]{mm} distance between the films. The films are tilted \unit[5]{$^\circ$} from the ion axis to reduce the influence of reflections on the measured pressure signal.}
\end{figure}

The experiments were carried out at the PHELIX laser facility at the GSI in Darmstadt, Germany. The PHELIX laser provides pulses of around \unit[150]{J} in \unit[500]{fs} at a central wavelength of \unit[1053]{nm}. The laser was focused by a \unit[45]{$^\circ$}, f=\unit[400]{mm} (F/1.4) off-axis parabolic mirror (OAP) onto a \unit[1]{\textmu m} thick polymer foil target. The proton beam was characterised by RCF-stacks and the proton cut-off energy reached between \unit[50]{MeV} and \unit[60]{MeV}, with an exponential decay typical for target-normal sheath acceleration (TNSA \cite{snavely_intense_2000}). The majority of shots was used to test the I-BEAT detector. Figure \ref{fig:exp-setup} shows the experimental setup.

\begin{figure} [ht]
  \begin{center}
  \begin{tabular}{ccc} 
  \begin{tikzpicture}
   \node[anchor=south west, inner sep=0](X) at %
   (0, 0){\includegraphics[width=.3\textwidth]{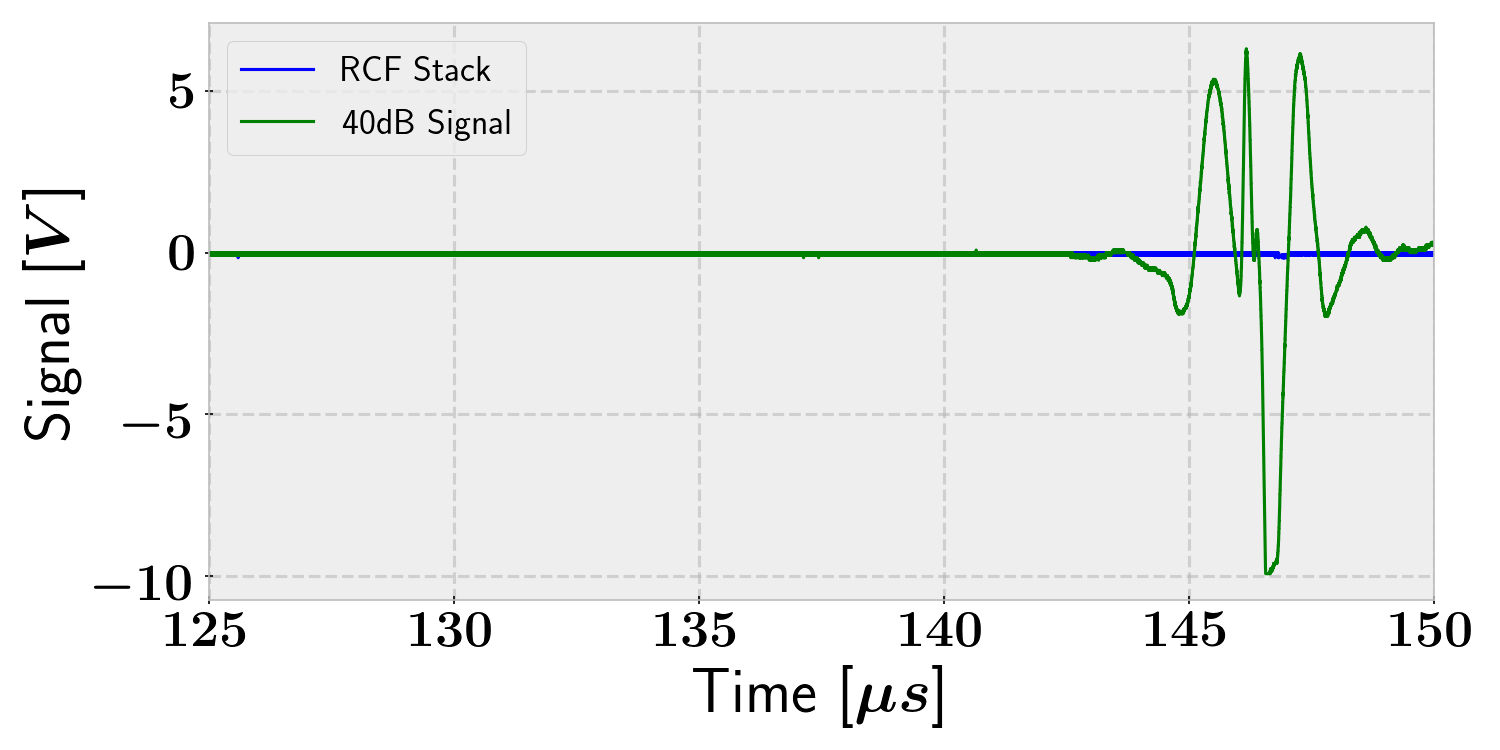}};%
   \begin{scope}[x={(X.south east)}, y={(X.north west)}]%
   \node[anchor=north west, black, fill=white](Y) at (0, 1){a)};%
   \end{scope}
  \end{tikzpicture} &
  \begin{tikzpicture}
   \node[anchor=south west, inner sep=0](X) at %
   (0, 0){\includegraphics[width=.3\textwidth]{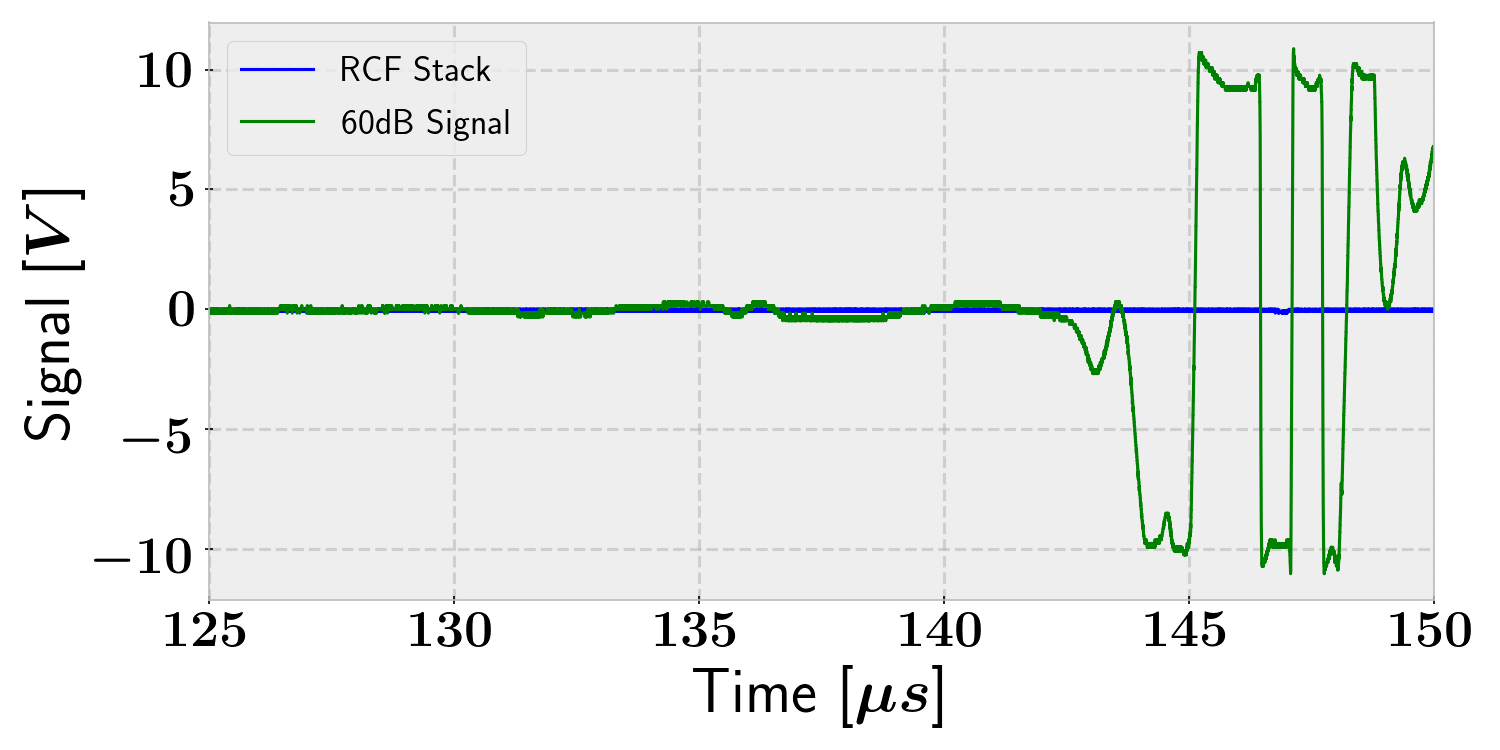}};%
   \begin{scope}[x={(X.south east)}, y={(X.north west)}]%
   \node[anchor=north west, black, fill=white](Y) at (0, 1){b)};%
   \end{scope}
  \end{tikzpicture} &
  \begin{tikzpicture}
   \node[anchor=south west, inner sep=0](X) at %
   (0, 0){\includegraphics[width=.3\textwidth]{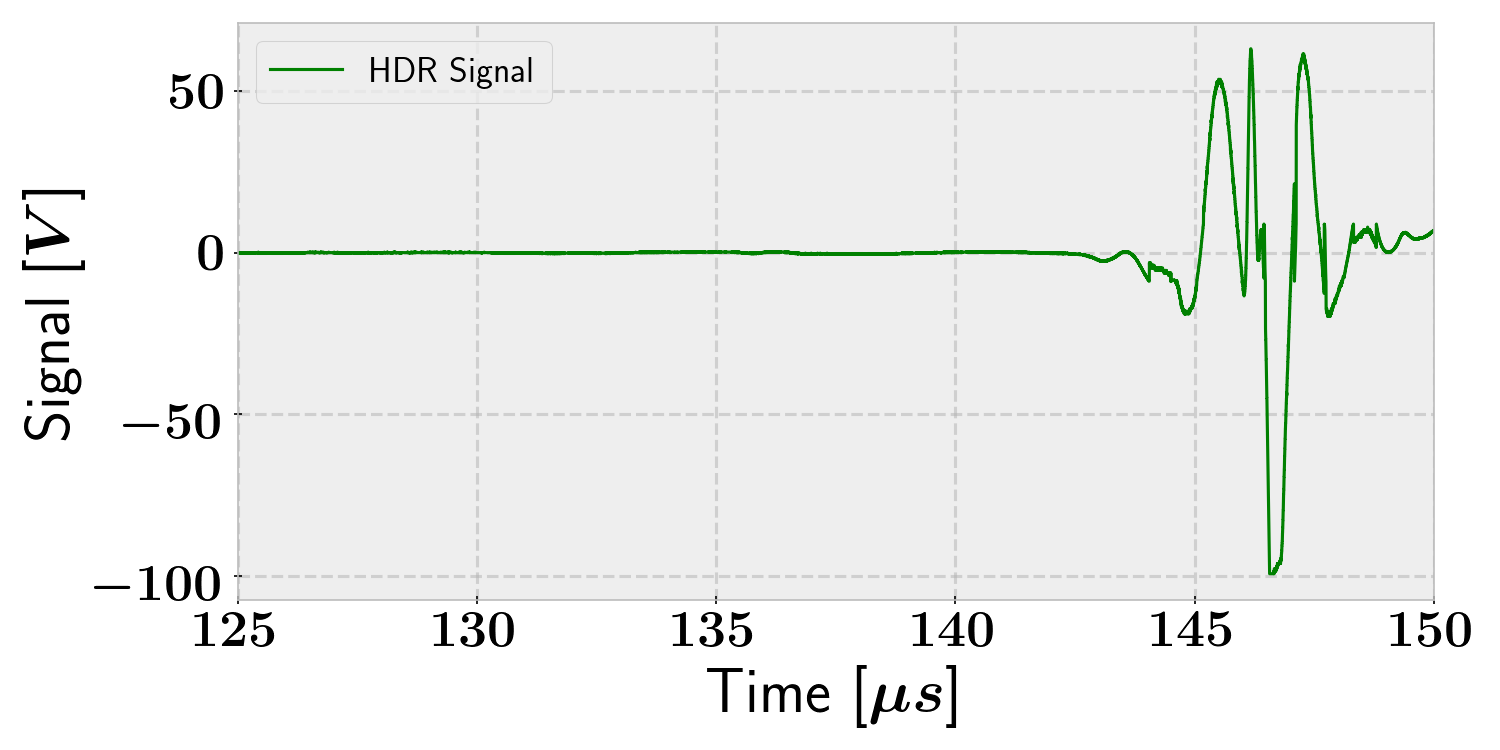}};%
   \begin{scope}[x={(X.south east)}, y={(X.north west)}]%
   \node[anchor=north west, black, fill=white](Y) at (0, 1){c)};%
   \end{scope}
  \end{tikzpicture}
  \end{tabular}
  \end{center}
  \caption[Axial Traces] 
  { \label{fig:ax-traces} 
The signal of the axial transducer in green. To illustrate that the signal is due to dose deposition in the detector, the trace recorded with an RCF stack blocking the entrance window is shown in blue. The two signals recorded with amplifications of \unit[40]{dB} (a) and \unit[60]{dB} (b) can be combined to a single trace (c) to increase the dynamic range. This introduces only minor artefacts at the stitching points between the traces.}
\end{figure}
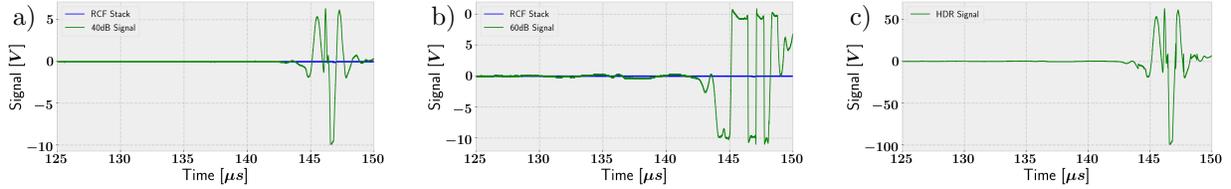

The signal recorded with the axial transducer has a high amplitude and is related to proton impact. We tested this by blocking the entrance in a number of shots. Occasionally, we did not use a target when firing the laser, in such shots we also observed no signal. Fig. \ref{fig:ax-traces} shows two recordings with protons using amplification settings of (a) \unit[40]{dB} and (b) \unit[60]{dB}. With the lower amplification the peak structure could be resolved without saturating the acquisition electronics. At higher amplification the detector saturated. However, the signal at earlier times could be resolved, which was before lost in the digitization of the acquisition. Combination of both signals by replacing the saturated parts of the highly amplified signal with the corresponding re-scaled parts of the less amplified signal allows for an increase in dynamic range of the signal from 8-bit to more than 10-bit. Fig. \ref{fig:ax-traces}c shows this combined trace and we will further analyse this signal. It has to be noted, that the visibility of the signal onset is still limited by the digitization of the acquisition electronics and not by the signal-to-noise ratio.
The signal evidences a strong modulation with a frequency of around \unit[1]{MHz}. This is the resonance frequency of the transducer (Olympus V303-SU) and is likely not related to the actual shape of the energy density (or dose) distribution along depth into the water. Ideally, this could be understood by correcting for the frequency dependent response of the transducer and its subsequent amplifier(s), but this has not been accomplished yet. Therefore, we will restrict our analysis to a qualitative comparison of the measured signal to the idealised signal that we expect from an exponentially decaying energy distribution that will cause a likewise exponentially decaying depth dose curve (\ref{sec:sim}). 

\begin{figure} [ht]
  \begin{center}
  \begin{tabular}{cc} 
  \begin{tikzpicture}
   \node[anchor=south west, inner sep=0](X) at %
   (0, 0){\includegraphics[width=.4\textwidth]{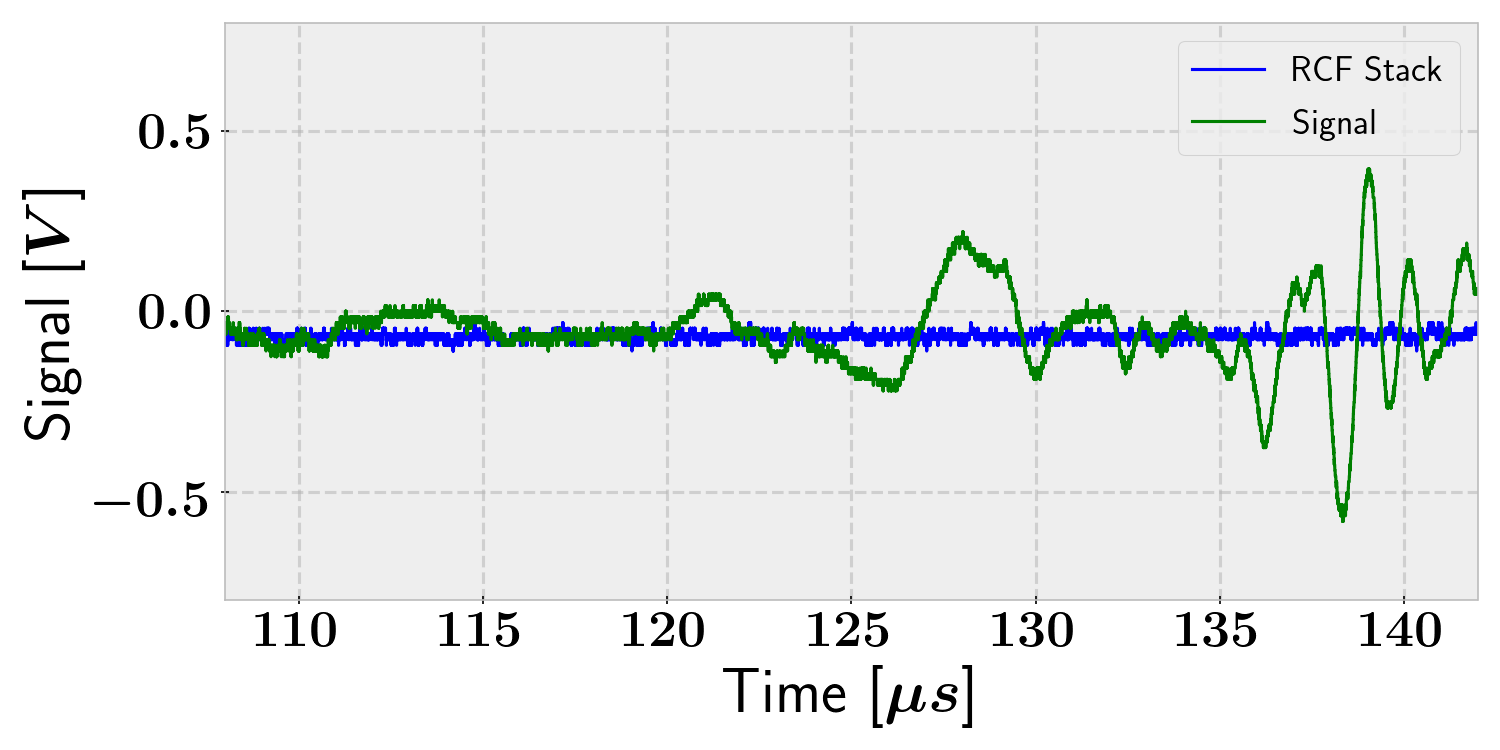}};%
   \begin{scope}[x={(X.south east)}, y={(X.north west)}]%
   \node[anchor=north west, black, fill=white](Y) at (0, 1){a)};%
   \end{scope}
  \end{tikzpicture} &
  \begin{tikzpicture}
   \node[anchor=south west, inner sep=0](X) at %
   (0, 0){\includegraphics[width=.2\textwidth, angle=90]{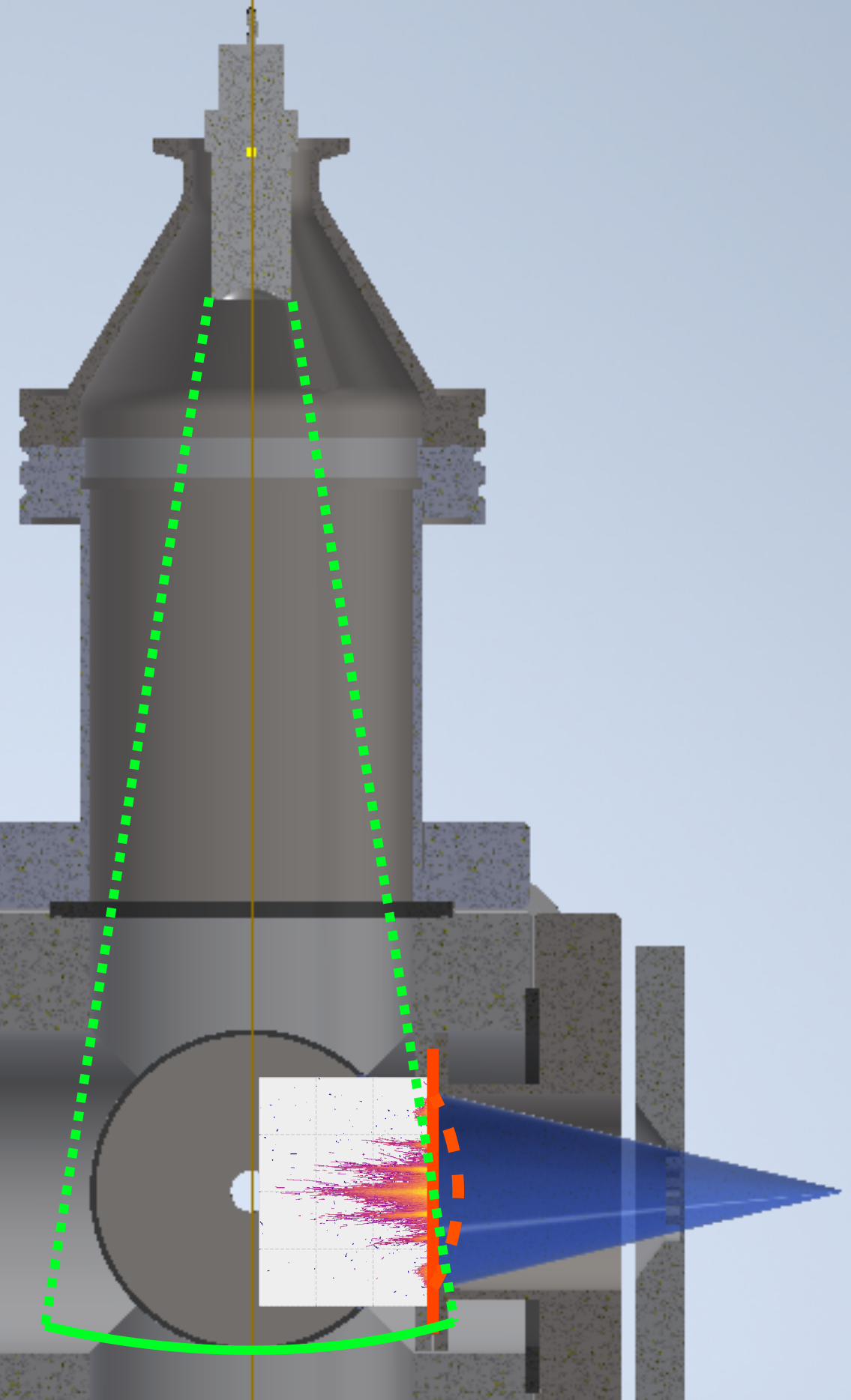}};%
   \begin{scope}[x={(X.south east)}, y={(X.north west)}]%
   \node[anchor=north west, black, fill=white](Y) at (0, 1){b)};%
   \end{scope}
  \end{tikzpicture}
  \end{tabular}
  \end{center}
  \caption[Lateral Trace] 
  { \label{fig:lat-trace} 
The trace of the lateral transducer (a), taken with \unit[60]{dB} amplification. For comparison, a shot with the entrance window blocked by an RCF stack is shown in blue. The particular detector geometry (b) can explain the lack of signal in the lateral transducer. The ion cone is indicated in blue and the field of view of the lateral transducer in dashed green with the solid green line corresponding to a signal runtime of \unit[140]{\textmu s}. The simulated energy deposition behind the entrance window (orange) is shown to give a scale for the penetration depth. When the entrance window deforms due to the pressure difference (from the solid to the dashed line), the volume of highest dose deposition is no longer in the field of view of the transducer.}
\end{figure}

Fig. \ref{fig:lat-trace}) shows a typical signal recorded with the lateral transducer. Modulations are visible as well, but reach much smaller amplitudes than for the axial transducer. This lack of signal can be explained by the detector geometry and the shape of the energy density distribution deposited by the protons in the water volume (Fig. 3b). Due to the exponentially decaying spectrum at the source, the highest energy deposition is located at the entrance window and it decays quickly with penetration depth. We noticed in the experiment that the entrance window was deformed and bulged out by several millimeters due to the pressure difference between the water reservoir and surrounding vacuum. Therefore, we concluded that there was no direct line-of-sight between the lateral transducer and the volume of highest energy deposition. Only the smaller energy deposition of the highest energetic protons reached into the field of view and resulted in the signal visible between \unit[125]{\textmu s} and \unit[130]{\textmu s}. However, this signal level is not sufficient to resolve lateral structures. The signal recorded between \unit[135]{\textmu s} and \unit[140]{\textmu s} is assumed to be the first reflection of the pressure wave at the detector walls.

For the axial transducer, the deformation of the entrance window leads to a temporal shift of the signal by a few microseconds, but no significant decrease in amplitude. 

\section{SIMULATIONS}
\label{sec:sim}

In thermo-acoustic approximation, the general solution to the wave equation reads
\begin{equation}
    \label{eq:pressure}
	p(\vec{r}, t) = \frac{\Gamma}{c_s^2}\frac{\partial}{\partial t} \int d^3 \vec{r}^{\, \prime} \frac{1}{|\vec{r}-\vec{r}^{\, \prime}|}H \left( \vec{r}^{\, \prime}, t-\frac{|\vec{r}-\vec{r}^{\, \prime}|}{c_s} \right)
\end{equation}
where $\Gamma$ and $c_s$ are the Gr\"uneisen-Parameter and speed of sound of water, $\vec{r}$ is the coordinate of the transducer, $t$ is time and $\vec{r'}$ is the coordinate of source point. We assume that the heating function $H$ is the product of a delta-function in time, because the proton bunch impact happens on time scales much shorter than the period of the observed pressure pulse. Then, we can calculate the pressure pulse at the position of the transducer for a given three-dimensional dose distribution.

\begin{figure} [ht]
  \begin{center}
  \begin{tabular}{cc} 
  \begin{tikzpicture}
   \node[anchor=south west, inner sep=0](X) at %
   (0, 0){\includegraphics[width=.35\textwidth]{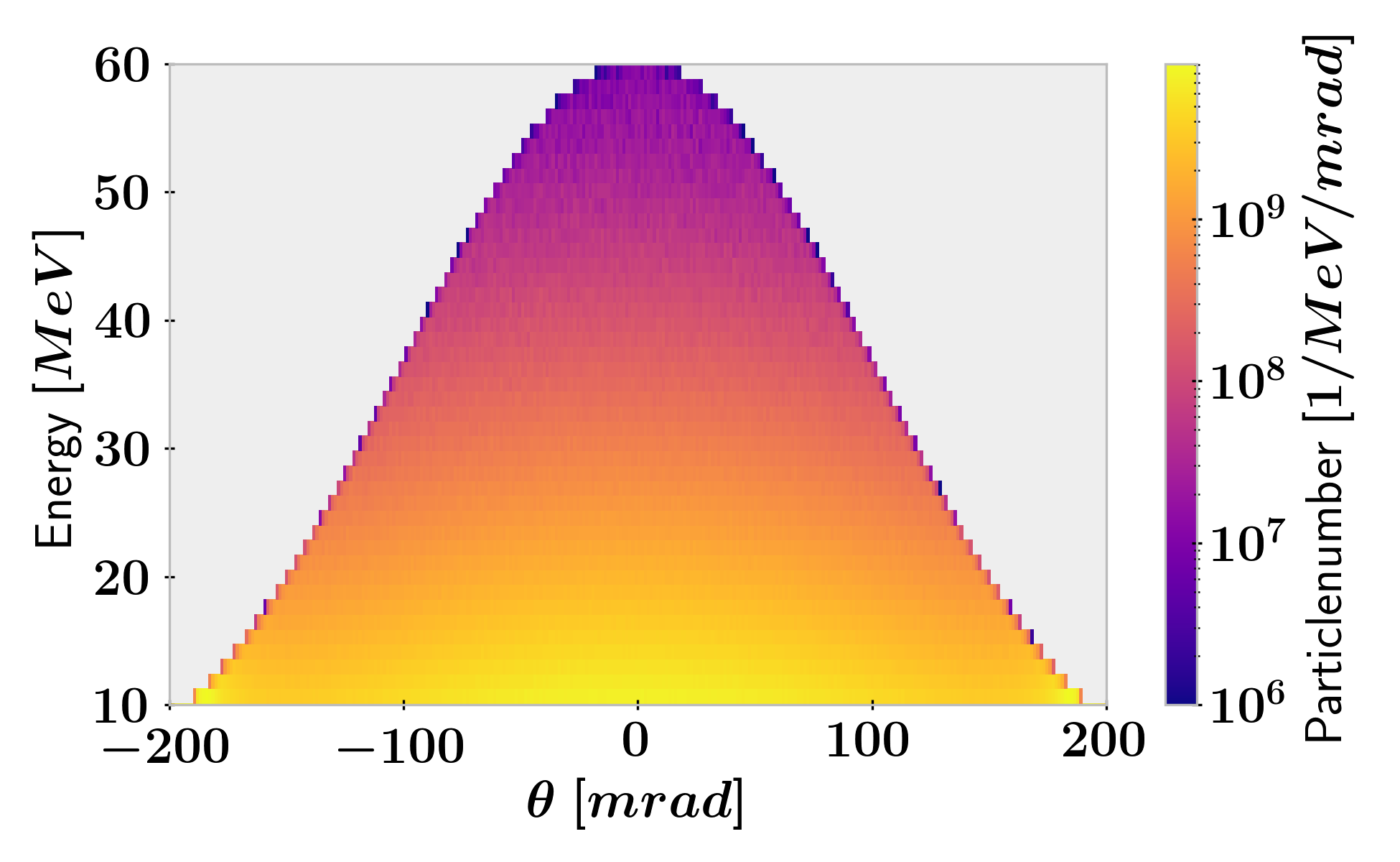}};%
   \begin{scope}[x={(X.south east)}, y={(X.north west)}]%
   \node[anchor=north west, black, fill=white](Y) at (-0.05, 1.05){a)};%
   \end{scope}
  \end{tikzpicture} &
  \begin{tikzpicture}
   \node[anchor=south west, inner sep=0](X) at %
   (0, 0){\includegraphics[width=.45\textwidth]{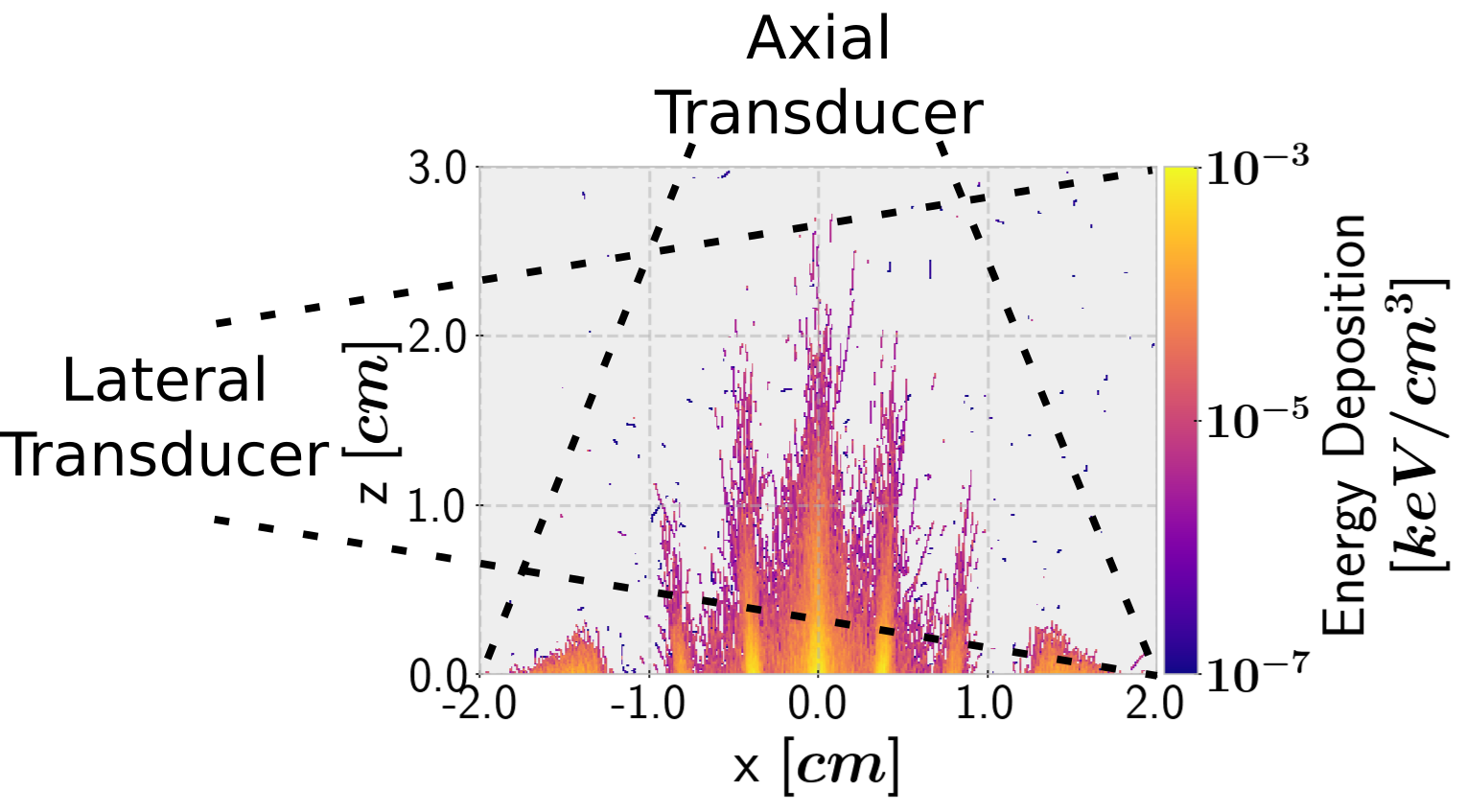}};%
   \begin{scope}[x={(X.south east)}, y={(X.north west)}]%
   \node[anchor=north west, black, fill=white](Y) at (0, 0.9){b)};%
   \end{scope}
  \end{tikzpicture}\\
  \begin{tikzpicture}
   \node[anchor=south west, inner sep=0](X) at %
   (0, 0){\includegraphics[width=.45\textwidth]{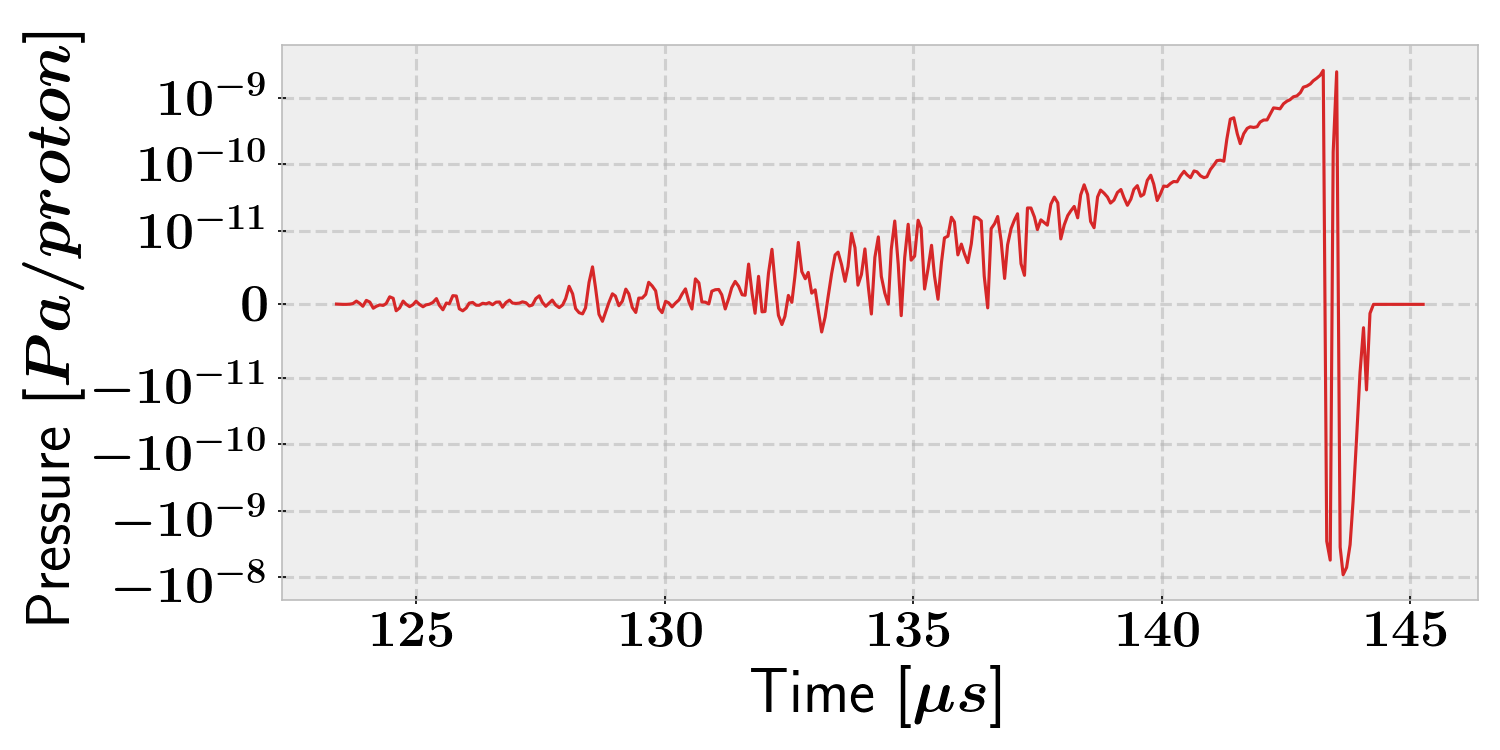}};%
   \begin{scope}[x={(X.south east)}, y={(X.north west)}]%
   \node[anchor=north west, black, fill=white](Y) at (0.2, 0.95){c)};%
   \end{scope}
  \end{tikzpicture} &
  \begin{tikzpicture}
   \node[anchor=south west, inner sep=0](X) at %
   (0, 0){\includegraphics[width=.45\textwidth]{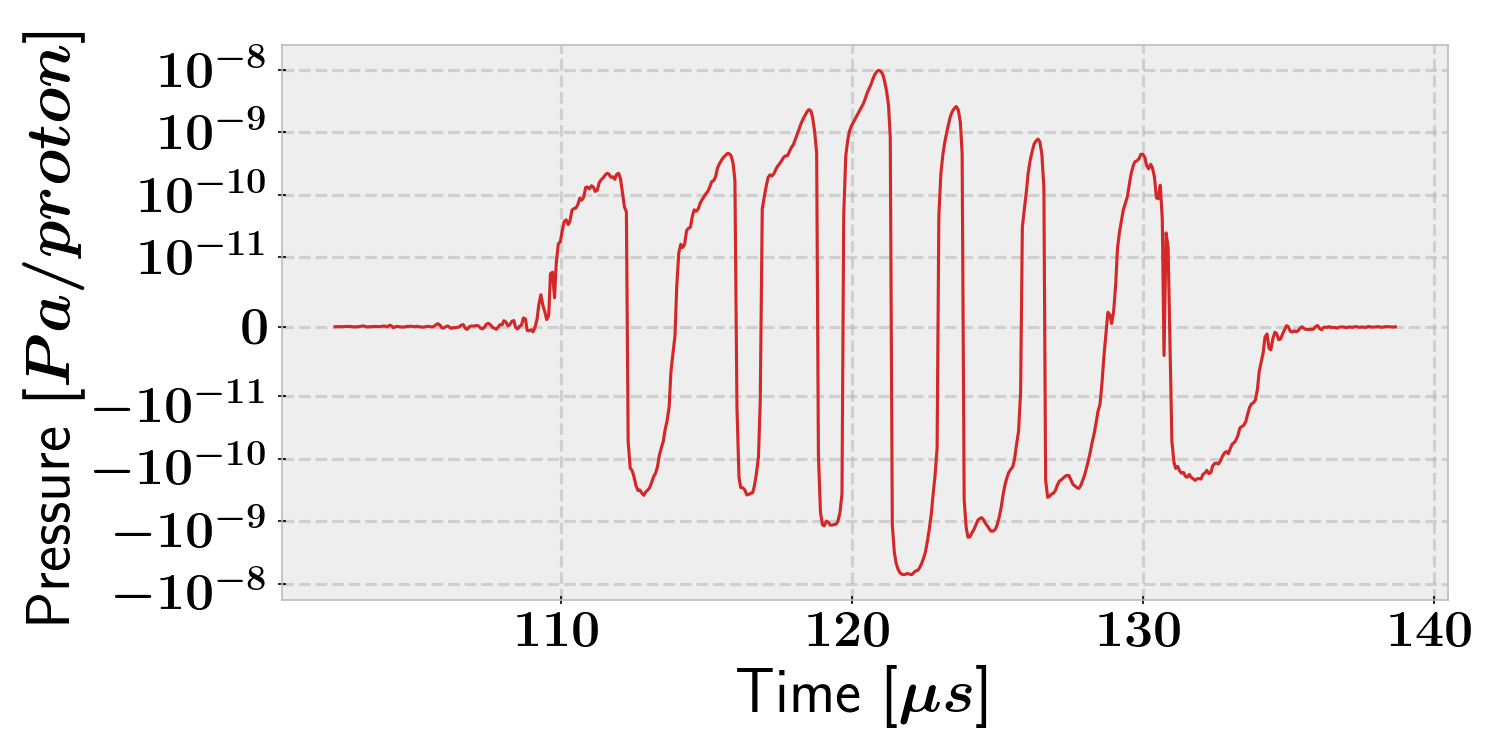}};%
   \begin{scope}[x={(X.south east)}, y={(X.north west)}]%
   \node[anchor=north west, black, fill=white](Y) at (0.2, 0.95){d)};%
   \end{scope}
  \end{tikzpicture}
  \end{tabular}
  \end{center}
  \caption[Simulation] 
  { \label{fig:simulation} 
Simulation setup and results. (a) The energy-angular distribution of protons used as input for the FLUKA simulation. The divergence angle is Gaussian distributed and the energy spectrum for each angle decays exponentially up to a cut-off. The cut-off energy follows in turn a Gaussian distribution over the divergence angle. (b) The energy deposition inside the water volume (including the entrance window) for a slice through the ion beam axis is shown normalized to a single proton. The direction and position of the transducers are indicated by the dashed lines. In the lower half the pressure at the axial (c) and lateral (d) transducer, calculated from the simulated energy density distribution, is shown.}
\end{figure}
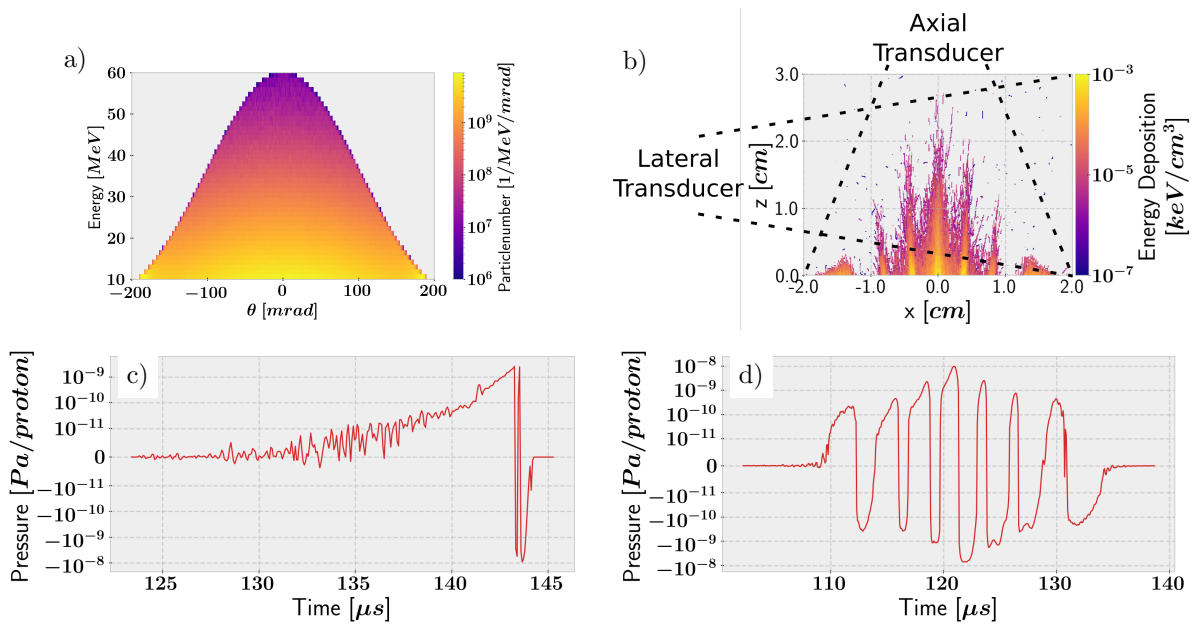

This dose distribution is obtained by simulations in FLUKA.\cite{ferrari2005fluka}\cite{bohlen2014fluka} Fig. \ref{fig:simulation})a shows the differential energy distribution as logarithmic color scale plot which was used as input for the ion bunch. The simulation modelled the propagation of the ion bunch through the Aluminum mask and the Kapton entrance window into the water reservoir. The deposited energy density distribution shows clearly the lateral modulation introduced by the mask (fig. \ref{fig:simulation}b). From the simulated energy deposition it is possible to solve the integral in Eq. \ref{eq:pressure} numerically for a given transducer position. Figure \ref{fig:simulation}c shows this acoustic trace for the axial transducer. The pressure rises over more than 4 orders of magnitude in a time window of about \unit[20]{\textmu s}, consistent with a penetration depth of \unit[30]{mm} expected for \unit[60]{MeV} protons. At the vacuum interface, the trace is abruptly truncated, whereas in reality we would expect a time-mirrored and inverted contribution from the reflection at this interface (which we did not include in our calculations yet). This high dynamic range of the expected pressure signal and the comparably slow rise time poses a particular challenge for detecting the full range of the signal. For completeness, Fig. \ref{fig:simulation}d shows the expected trace in the lateral transducer if the entrance window was not bulged out.

\section{DISCUSSION}
\label{sec:dis}

\begin{figure} [ht]
  \begin{center}
  \begin{tabular}{cc} 
  \includegraphics[width=.5\textwidth]{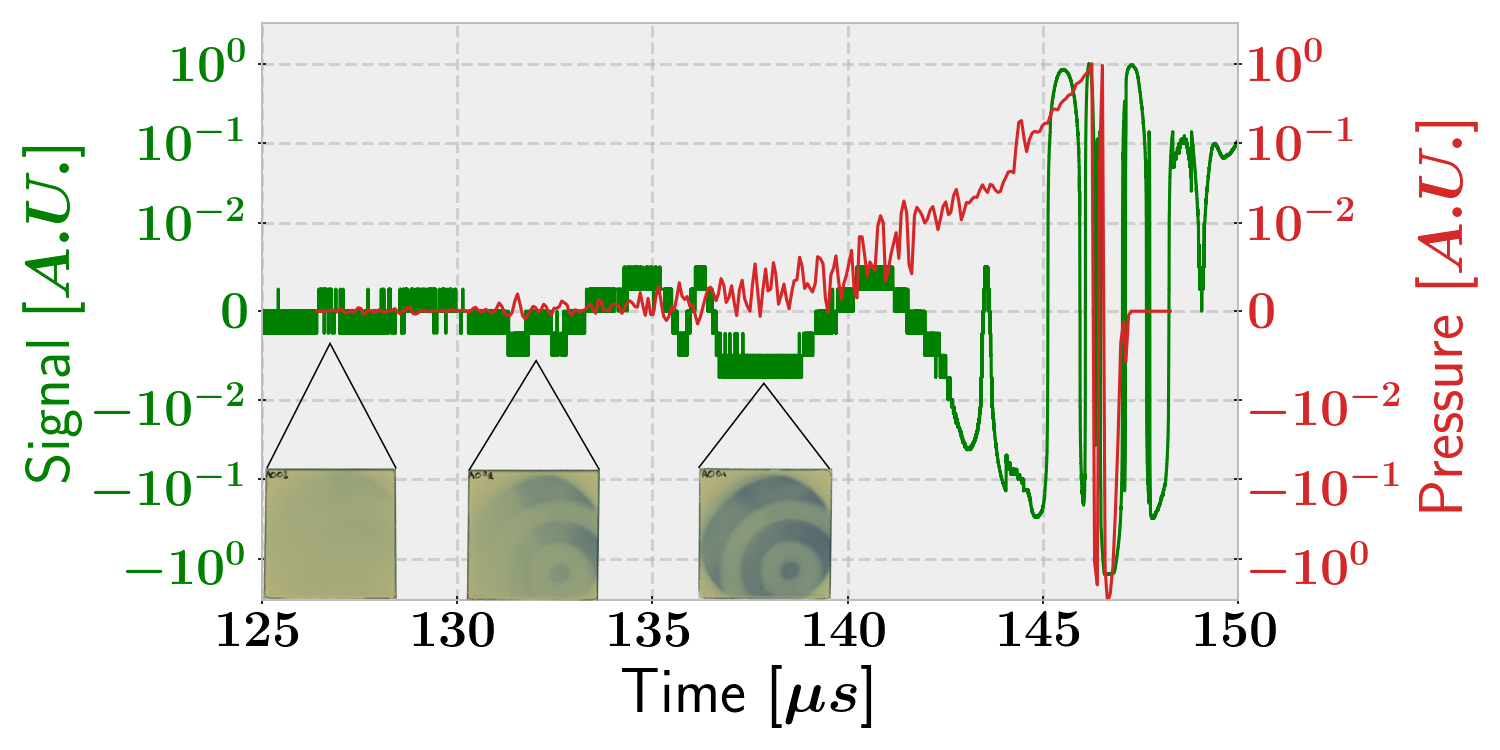}
  \end{tabular}
  \end{center}
  \caption[Comparison of Simulation and Experiment] 
  { \label{fig:comp} 
Comparison between experimentally measured (unfiltered) and simulated acoustic trace. The simulated pressure curve is shifted by \unit[3]{\textmu s} to compensate the deformation of the entrance window. Also shown are scans of the RCFs that were inserted in the reservoir at the indicated positions (corrected for the window shift as well)}
\end{figure}

When comparing the simulated pressure curve for the axial transducer (fig. \ref{fig:simulation}c) with the measured signal (fig. \ref{fig:ax-traces}c), it appears that the strongest contributions (originating from regions near the entrance window) are shifted by $\sim$ \unit[3]{\textmu s}. This suggest a deformation of the entrance window by \unit[4.5]{mm}, which is plausible when compared to a visible inspection during the experiment. We therefore correct for this shift. Fig. \ref{fig:comp} then directly compares the shifted simulated acoustic trace with the measured one (both normalized to a maximum of 1). It is obvious that there is little similarity between the two traces. This is because we here compare an ideal simulated waveform (red) with a waveform that was measured with a frequency dependent transducer and amplifier without correcting for the complex spatial and impulse response. It is very likely that both the switched polarity as well as the strong modulation are at least partially caused by this response. It is also worthwhile noting that the measured trace expresses strong modulations at the positions where RCFs were inserted.
Regardless of this, the temporal extent of the two signals are indeed comparable. Therefore, it is fair to approach some quantitative estimates. A signal is visible at \unit[132]{\textmu s}, that is approximately \unit[15]{\textmu s} prior to the window signal and hence \unit[22.5]{mm} into the water, which corresponds to a proton energy of \unit[50]{MeV}. At that depth, the gafchromic film is also notably darkened. At the position of the first film, \unit[14.5]{mm} into the water (\unit[39]{MeV}), the signal is stronger, as expected for an energy distribution that is decaying towards higher energies.

Regarding the lateral transducer, the simulations predicted a signal strength in the same order of magnitude as the axial case (\ref{fig:simulation}d). When considering the window shift of \unit[4.5]{mm}, which we derived from the measured trace shift, fig. \ref{fig:simulation}b explains qualitatively, that this shift can indeed be responsible for moving the strongest part of the source out of the observation cone of the lateral transducer. Hence, it is conceivable that this explains the lack of signal at the lateral transducer.

\section{CONCLUSION}

We have demonstrated the progress of the ionoacoustic approach to detect laser accelerated protons in the harsh environment close to the laser-plasma interaction without measures for energy selection. In particular, the ionoacoustic signal decoupled well from the EMP and no radiation damage to the detector was observed. The signal (in particular in lateral beam direction) was limited by the detector geometry. The axial transducer serves well for recording acoustic traces, showing a correlation with the signal on the RCFs. However, the currently limiting factor is the dynamic range of the recorded signal. We judge, that an increase to 14 or 16 bit will be sufficient for a significant improvement of the quality of the measured traces.
Detailed simulations that include the complete geometry, most importantly the RCF-films within the beam path, must be performed in future studies. This geometry affects not only the energy density distribution but also the propagation of the emerging sound wave and hence the shape of the acoustic trace. In addition, the spatial and impulse response of the transducer-amplifier system must be quantified, ideally via appropriate calibration sources.
We believe that the I-BEAT approach remains promising, in particular because it promises higher resolution and retrieval of lateral bunch information with further development and increasing proton energies. Already in its current state and with the current dynamic range, the signal of the axial detector would be suitable as a qualitative online diagnostic for source optimization at higher repetition rates.

\acknowledgments
The results presented here are based on the experiment P205, which was performed at the PHELIX facility at the GSI Helmholtzzentrum fuer Schwerionenforschung, Darmstadt (Germany) in the frame of FAIR Phase-0. 
This work was supported by the BMBF under project 05P18WMFA1. SG acknowledges the support of the German Research Foundation (DFG) within the Research Training Group GRK 2274.

\bibliography{report} 
\bibliographystyle{spiebib} 

\end{document}